\documentclass[11pt,a4paper,twoside,openright]{article}

\usepackage[english]{babel}
\usepackage[dvips]{graphicx}
\usepackage[colorlinks]{hyperref}
\usepackage[numbers]{natbib}
\usepackage{amsmath,amssymb,amsthm,verbatim,amscd,mathrsfs,amsfonts,latexsym,epic,hypernat} 
\usepackage{psfrag,epsfig}
\usepackage[usenames]{color}
\usepackage[utf8]{inputenc}
\usepackage[ruled,vlined]{algorithm2e}
\usepackage{multirow}
\usepackage{pdflscape}  

\usepackage{tikz}
\usetikzlibrary{arrows,chains,matrix,positioning,scopes}

\makeatletter
\tikzset{join/.code=\tikzset{after node path={%
\ifx\tikzchainprevious\pgfutil@empty\else(\tikzchainprevious)%
edge[every join]#1(\tikzchaincurrent)\fi}}}
\makeatother
\tikzset{>=stealth',every on chain/.append style={join},
         every join/.style={->}}
\tikzstyle{labeled}=[execute at begin node=$\scriptstyle,
   execute at end node=$]

\newcommand{\esp}{\mathbb{E}}

\newcommand{\A}{\mathcal{A}}
\newcommand{\Z}{\mathcal{Z}}

\newcommand{\med}{\mathop{\text{median}}}
\newcommand{\Ins}{\text{Ins}}
\newcommand{\Hom}{\text{Hom}}

\newcommand{\asso}{\to}

\newcommand{\red}[1]{{\color{red}{#1}}}

\title{A time warping approach to multiple sequence alignment}
\author{Ana Arribas-Gil$^1$ and Catherine Matias$^2$}

\begin{document}
\thispagestyle{empty}

\maketitle

\begin{center}
1.  Departamento  de  Estad\'istica,   Universidad  Carlos  III  de Madrid,  C/ Madrid, 126 - 28903 Getafe,  Spain.	   
E-mail: ana.arribas@uc3m.es \\
2. Sorbonne Universités, Université Pierre  et Marie Curie, Université Paris
Diderot, Centre National de la Recherche Scientifique, 
Laboratoire de  Probabilités   et    Modèles   Aléatoires, 4 place Jussieu, 75252 PARIS Cedex 05, FRANCE. 
E-mail:  catherine.matias@math.cnrs.fr
\end{center}

\begin{abstract}
We propose  an approach for  multiple sequence alignment (MSA) derived from
the  dynamic  time  warping  viewpoint and  recent
techniques of curve synchronization developed in the context of
functional data  analysis. Starting  from pairwise  alignments of all  the sequences
(viewed as paths in a certain  space), we construct a median path that
represents the MSA we are looking for. We  establish a proof  of concept  that our
method could be an interesting  ingredient to include into refined MSA
techniques. We present  a simple synthetic experiment as  well as the
study of a benchmark dataset,  together with comparisons with 2 widely
used MSA softwares. 
\end{abstract}



\section{Introduction}    

Multiple sequence alignment (MSA) is one of the most fundamental tasks in bioinformatics. While there are many attempts to
handle comparative sequence analyses without relying on MSA, it still represents a starting point for most evolutionary
biology methods.  
Pairwise sequence alignment has been conceptualized as early as the 1970's, starting with global alignments that attempt to align
entire sequences~\citep{Needleman1970} and then introducing a decade later local alignments that focus on the
identification of subsequences sharing high similarity~\citep{Smith1981}. The standard computational formulation of both tasks is to
maximize a scoring function obtained as the sum of the score for each aligned pair of residues (nucleotides or amino
acids, the highest scores being attributed to pairs of residues with highest similarity), minus some gaps penalties. Since these seminal works, an abundant literature has flourished exploring this topic in many different directions,
from the pairwise  problem to the more complex task of aligning more than 3 sequences~\citep[one of the very first
attempts appearing in][]{Lipman89},  from exact solutions that scale exponentially with
sequence lengths to faster heuristic approaches used in the most common tools, and from the scoring formulation of the alignment
problem that requires to choose the scoring parameters to probabilistic formulations in which those parameters are estimated~\citep{DEKM,AGM06}. However, manually refined alignments continue to be superior to purely automated methods and there is a continuous effort to improve the
accuracy of MSA tools~\citep{Edgar}. We refer the reader to the reviews~\citep{Edgar,Wallace,Kumar,Notredame} for more details
on MSA. \\

Dynamic time warping (DTW) is a general version of the dynamic programing algorithm that solves exactly the 
pairwise biological sequence alignment problem. It
is a  well-known and general technique to find an optimal alignment between two given (time-dependent)
sequences. 
In time series analysis, DTW is used for constructing an optimal alignment of two sequences with
possible different lengths  by stretching or contracting time intervals~\citep{Keogh2005}. 
In functional data analysis, the time warping approach consists in modeling a set of curves exhibiting time and
amplitude variation with respect to a common continuous process~\citep{Liu_Muller}. Thus, time warping techniques are used in many different
areas concerned by sequence or curve comparisons, one of its most famous successes being on human-speech
recognition~\citep{Kruskal_TW}. 

Here, we propose a simple and fast procedure for MSA, inspired from recent techniques of curve synchronization developed in the context of
functional data analysis~\citep{Tang_Muller,Ana_Muller}. In this setup, one often observes a set of curves which are modeled
as the composition of an 'amplitude process' governing their common behavior, and a 'warping process' inducing time
distortion among the individuals. Specifically, $y_i(t)=x_i\circ h_i(t)$, $t\in[a,b]$, $i=1,\dots,K$, are observed, with
$x_i$ being i.i.d. realisations of the amplitude process $X$, and $h_i$ strictly monotone functions such that  $h_i(a)=a$ and $h_i(b)=b$ being i.i.d. realisations of the warping process $H$. Aligning pairs of curves (that is eliminating time variation, which comes to estimating the warping functions $h_i$) is a first step before
estimating the common amplitude process. These authors proposed to first estimate pairwise warping functions between all
possible trajectories pairs which are then used to create estimators of the underlying individual warping
functions in a second step. Sample means or more robust median-based estimates  come into play to solve
this second step. This procedure is an alternative to the widely used approach of template registration, that consists in aligning every observed curve to some given template, which should be a good estimate of the mean amplitude process. The drawback of this methodology is that it heavily relies on the choice of the template, which is not straightforward. Now, in the MSA context, the warping process is the insertion-deletion (or indel) process
that  stretches or contracts the initial sequence, while the amplitude process is the substitution process that modifies
the value of the sequence base. The equivalent of template registration in the MSA context would be the alignment of every sequence to some estimate of the ancestral sequence, which is, of course, not available. However, exploiting the previous ideas, we show how pairwise alignments can be combined with a simple
median-based approach to obtain an estimate of the multiple alignment of the sequences.

Our aim is to  establish a proof of concept that  our new method could
be   an   interesting  ingredient   to   include   into  refined   MSA
techniques. Indeed, the method is able  to align a large number $K$ of
sequences (that are assumed to share a common ancestor) in a 
quite simple and fast manner, although a bit rough w.r.t. accuracy. 
We would like to stress that we do not claim to be competitive with 
actual  aligners. Let us recall that there already exist many competitors to solve the MSA problem whose respective
performances have been largely explored~\citep[see for e.g][and the references therein]{Pais2014}. 
Here, we would rather like to point out to recent developments from curve synchronization that could
open the way to new improvements in MSA. While we do not pretend to propose a most accurate method, it is important to
note that our approach could  be used  as  a starting  point in an 
iterative refinement strategy~\citep{Gotoh}. Those strategies, included in many widely used tools such as 
\texttt{ProbCons}~\citep{ProbCons}, \texttt{MUSCLE}~\citep{Muscle}, \texttt{MAFFT}~\citep{Mafft} or
\texttt{MUMMALS}~\citep{Mummals}, mostly consist in repeatedly dividing the set of aligned sequences
into  two random  groups  and  realign  those  groups by optimizing an objective function. Thus, our simple and fast
procedure could be combined with similar refinement strategies that would improve its efficiency.  
An advantage of our method is  that  it only  uses  pairwise comparison but (unlike progressive
aligners) is not sensitive to any order on which we consider the sequence pairs. Indeed, progressive aligners rely on a
guiding tree  and progressively  build $K-1$ pairwise  alignments from
the union of the initial set of $K$ sequences with consensus
sequences built at the internal nodes of the tree. As such, the consensus sequences heavily
depend on the guiding tree and the sequences ordering which impacts the resulting MSA. 
Moreover, progressive aligners  are known to tend
to propagate errors appearing at the early stages of the method. For instance, \texttt{ClustalW}~\citep{ClustalW} is criticized
for being responsible of the heuristic rule 'once a gap, always a gap' that makes errors in pairwise alignments to
propagate through MSA. The method proposed here is not sensitive to that issue because a gap in an alignment pair will
not necessarily be selected for the MSA.

The manuscript is organized as follows. Section~\ref{sec:model} first recalls concepts coming from the pairwise alignment and
time warping problems (Section~\ref{sec:rappels}), sets our framework (Section~\ref{sec:MSA}) and describes our
procedure (Section~\ref{sec:estim}). Then Section~\ref{sec:results} presents our experiments, starting with synthetic
datasets (Section~\ref{sec:synthetic}), then relying on a benchmark dataset (Section~\ref{sec:balibase}), namely the
Balibase dataset~\citep{Balibase} and concludes with possible extensions of our approach (Section~\ref{sec:ext}).


\section{Framework  and procedure}
\label{sec:model}
In what follows, we first introduce some ideas around pairwise alignment and the time warping problem. We then explain
how to use these concepts in the context of multiple sequence alignment.

\subsection{Pairwise alignment and time warping}
\label{sec:rappels}
\begin{figure}[!h]
\centering

\begin{tikzpicture}[scale=0.5] 
\draw[->] (0,0)--(0,8.5);
\draw[->] (0,0)--(8.5,0);
\draw [thin, gray,dotted] (0,0) grid (8,8);

\draw[loosely dashed] (8,0)--(8,6);
\draw[loosely dashed] (0,6)--(8,6);
\node (nm) at (9.5,6.5) {$(n,m)$}; 
\node (bullet) at (8,6){$\bullet$}; 

\draw [very thin](0,0) --(8,8);

\node (X1) at (0.5,-0.5) {$X_1$};   
\node (Xn) at (7.5,-0.5) {$X_n$};   
\node (Y1) at (-0.5,0.5) {$Y_1$};   
\node (Yn) at (-0.5,5.5) {$Y_n$};   

\draw[very thick, blue] (0,0)--(1,0)--(3,2)--(4,2)--(6,4)--(6,5)--(7,6)--(8,6);
\draw[dashed,red] (0,0)--(0,1)--(2,3)--(2,4)--(4,6)--(5,6)--(6,7)--(6,8);

\end{tikzpicture}

\caption{Graphical   representation  of   an  alignment   between  two
  sequences $X_1\cdots X_n$ and $Y_1\cdots Y_m$.  The blue and thick line corresponds to a
  càd-làg function $\phi_{X,Y}$ whose generalized inverse $\phi_{Y,X}= \phi_{X,Y}^{-1}$ is shown in dashed and red.}
\label{fig:pairwise_align}
\end{figure}
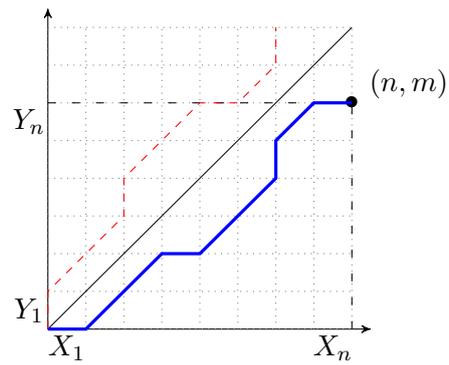

For any two sequences  $X=X_1\cdots X_n$ and $Y=Y_1\cdots Y_m$ with values in a finite
alphabet $\A$,  any 
(global) pairwise alignment of $X$ and $Y$ corresponds to an increasing path 
in the grid $[0,n]\times [0,m]$ composed by three elementary steps $\{(1,0),
(0,1), (1,1)\}$, as shown in Figure~\ref{fig:pairwise_align}. Note that for biological reasons, such path is often restricted
to never contain two consecutive steps in $\{(1,0),(0,1)\}$ (a gap in one sequence may not be followed by a gap in the
other sequence). We do not use that constraint in what follows but a post-processing step could be applied to any
alignment in order to satisfy it (the simplest way is then to replace those 2 successive gaps with a (mis)-match move; from a graphical point of view a
horizontal+vertical or vertical+horizontal move is replaced by a diagonal one). 
For notational convenience, we extend the path to  $[-\epsilon,n]\times [-\epsilon,m]$ for some small 
$\epsilon >0$. Now,  any such path may be viewed as an increasing function $\phi_{X,Y} :[-\epsilon,n] \to
[-\epsilon,m]$  with the convention that it is càd-làg (continuous on the
right, limit on  the left). By letting $\phi(u-)$ denote  the left limit of $\phi$ at $u$, we moreover impose that
$\phi_{X,Y}(0-)=0$  and  $\phi_{X,Y}(n)=m$. Note that
we can define a unique generalized inverse function $\phi_{Y,X}:=\phi_{X,Y}^{-1}: [-\epsilon,m] \to
[-\epsilon,n]$ constrained  to be càd-làg and such  that $\phi_{X,Y}\circ \phi_{X,Y}^{-1}=
\phi_{X,Y}^{-1} \circ \phi_{X,Y} = Id$ (the identity function).  From a graphical point of view, the
path corresponding to  $\phi^{-1}_{X,Y}$ is  obtained as  the symmetric  of the path corresponding to  $\phi_{X,Y}$ with
respect to the diagonal line $y=x$ (see Figure~\ref{fig:pairwise_align}).

Functions $\phi_{X,Y}, \phi_{X,Y}^{-1}$ may be viewed as time warping functions that describe the homology (and thus
also the indels) between
sequences $X,Y$ obtained from their alignment. Let us first explain this idea on a simple example. Consider an 
alignment between $X=ACAGTAGT$ and $Y=CTTAAG$ given as follows
\[
\begin{tabular}{ccccccccc}
A & C & A & G &T &A & - & G & T\\
- & C &T & -& T& A& A& G & - 
\end{tabular}
\]
(This alignment corresponds exactly to the thick and blue path depicted in Figure~\ref{fig:pairwise_align}).
In this alignment, any character from one sequence may be associated to a unique character from the other sequence as
follows: when the position corresponds to a \emph{match} or a \emph{mismatch}, the character is associated to the (mis)-matching character (in the other
sequence) while at positions corresponding to indels, we (arbitrarily) associate it to the character in the previous (mis)-matching position in
the other sequence. In order to deal with the case where no such previous (mis)-matching position exists (namely when the alignment begins with
an indel), we introduce extra artificial characters ($X_0,Y_0$) at the beginning of each sequence, these two artificial characters being
aligned with each other. In other words, we consider the alignment  
\[
\begin{tabular}{cccccccccc}
$X_0$ & A & C & A & G &T &A & - & G & T\\
$Y_0$ & - & C &T & -& T& A& A& G & - 
\end{tabular}
\]
The bare alignment (the alignment without specification of nucleotides) of these two sequences is as follows 
\[
\begin{tabular}{cccccccccc}
$\bullet $& B & B & B & B &B &B & - & B & B\\
$\bullet$ & - & B &B & -& B& B& B& B & - 
\end{tabular}
\]
Here $B$ stands for \emph{base} and we do not specify to which character it corresponds. 
The bare alignment naturally appears to separate the indel process from the substitution
process~\citep[see for e.g.][]{TKF,Ana_Metzler}. 
Given the sequences, the knowledge of the bare alignment is sufficient to recover the corresponding full alignment. 
In this context, the time warping process is exactly the indel process that stretches or contracts an initial sequence, while
the amplitude process is the substitution process that modifies the value of the sequence. We focus on
the time warping process encoded in the bare alignment. 

Let $\asso$ denote this (asymmetric) relation induced by the bare alignment. In our example, we have that $X_1$ is
associated to the artificial character $Y_0$ and we denote $X_1\asso Y_0$. In the same way, 
$X_2 \asso Y_1$, $X_3\asso Y_2$, $X_4\asso Y_2$, $X_5\asso Y_3$,
$X_6\asso Y_4$, $X_7\asso Y_6$ and $X_8 \asso Y_6$.  
Note that no position of $X$ is associated to $Y_5$ (this is because of a discontinuity of the function $\phi_{X,Y}$ at
$u=6$ where $\phi_{X,Y}(6-)=4$ and $\phi_{X,Y}(6)=5$).   More importantly, these associations are exactly encoded in  the mapping
$\phi_{X,Y}$ as we have $X_u \asso Y_v $ if and only if $ \phi_{X,Y} (u-)=v$.

More generally, for any real value $u\in [0,n]$, letting $\lceil u \rceil$ denote the smallest integer larger or equal to $u$, we set
$X(u):=X_{\lceil u\rceil}$ (and similarly for sequence $Y$). From the (bare) pairwise alignment of $X,Y$, we obtain a time warping function
$\phi_{X,Y}$ such that this alignment is entirely described by the association $X(u) \asso Y\circ \phi_{X,Y}(u)$. 
In what follows, we denote this functional association by $X \asso Y\circ \phi_{X,Y}$. Note that we equivalently have $Y \asso
X\circ \phi_{X,Y}^{-1}= X\circ \phi_{Y,X}$. 

We also mention that when the pairwise alignment between $X,Y$ is extracted from a multiple
sequence alignment containing  at least 3 sequences,  say $X,Y,Z$, the
above  relation  $\asso$  should  be associative  in  the  sense  that
whenever $X \asso  Y\circ \phi$ and $Y \asso Z\circ  \phi'$, we should
also have  $X \asso Z \circ  \phi' \circ \phi$. This  property will be
used in the next section.

\subsection{Multiple sequence alignment}
\label{sec:MSA}
In this section, we consider a set of sequences $S^1,\ldots, S^K$ with values in a finite
alphabet $\A$  and respective lengths $n_1,\ldots, n_K$  and we assume
that  they all share some latent  ancestor  sequence $A$ with values in $\A$
and length $N$. 

A multiple sequence alignment of the set of sequences $S^1,\ldots, S^K$ is given by the knowledge of \emph{homologous}
positions as well as \emph{inserted} positions. To fix the ideas, let us consider for example the following multiple alignment of 3 sequences
\[
\begin{tabular}{cccccccccc}
&  & (H) & (H) & (H)  & & &  & (H) & \\
$X_0$ & A & C & A & G &T &A & - & - & T\\
$Y_0$ & - & C &T & -& C& C& A& G & -  \\
$Z_0$ & T & C & A & C & - & - & - & C & T\\
\end{tabular}
\]
where the first line indicates homologous positions (H) while the other are inserted positions. 
Homologous positions describe the characters that derived from the
ancestor sequence, so that there are exactly $N$ homologous positions in the alignment (whenever an ancestral position was deleted in all
the sequences, this can not be reconstructed and such position would not appear in our reconstructed ancestral
sequence). For each homologous position, there is at most one character in sequence $S^i$ that is associated to it. This
means that homologous columns in the multiple sequence alignment may contain matches, mismatches and even gaps (when the
corresponding ancestral position has been deleted in that particular sequence). Between two consecutive homologous
positions, each sequence might have a certain number of characters that are inserted. These characters do not derive from an ancestral
position.  Note that these insert runs are not aligned in the sense that the choice of how to put the letters in the
insert regions is arbitrary and most MSA implementations simply left-justify insert regions. Now, given the set of
sequences, this multiple alignment may be completely encoded through the knowledge of the homologous positions in the
sequences (see Section~\ref{sec:algo}). 

Our goal is to estimate the alignment of each $S^i$ to
the ancestor $A$  and thus  the
global alignment of the set of sequences $S^1,\ldots, S^K$, by relying
on the set of pairwise alignments of each $S^i$ to all
the other sequences $S^1,\ldots,S^K$. To do this, we will implicitly assume that
a) the multiple sequence alignment of  all the sequences $S^i$ is well
approximated by the
alignment we would obtain from all the sequences $\{A, S^i ;   i= 1,\dots , K\}$;
b) all the  pairwise alignments of $S^i, S^j$  are good approximations
to the extracted pair alignments from the multiple alignment
of all sequences $\{A, S^i ;   i= 1,\dots , K\}$. \\

First, any sequence $S^i$ is derived from the common ancestor sequence $A$
through an evolutionary process that can be encoded in the alignment of these two sequences. This alignment induces a 
time warping process $\phi_i := \phi_{S^i,A} : [-\epsilon,n_i] \to [-\epsilon,N]$ such that we have  $S^i \asso A\circ \phi_i$. 
For the moment, we assume that the warping functions $\phi_i$'s are
i.i.d. realisations of the same random process $\Phi$. 
For identifiability  reasons, we constrain  $\Phi$ such
that $\esp(\Phi)= Id$ (see for example \citep{Ana_Muller}). Moreover, note that assuming the $\phi_i$'s are
i.i.d. realisations  of the same  process boils down to  considering a
star tree leading from the  ancestor sequence to any observed sequence
$S^i$. In the case of an ultra-metric tree, we would obtain non independent but identically distributed
  realisations. We will discuss later how to handle more realistic setups.

Since  the ancestral sequence $A$ is  not observed, the idea
of \citep{Ana_Muller} is to estimate the warping function $\phi_i$ from
all the pairwise alignments of $S^i$ with the other sequences. 
More precisely, when considering  the pairwise alignment of $S^i$ with
$S^j$,  we  obtain   a  warping  function  $g_{ji}=\phi_j^{-1}\circ  \phi_i$
satisfying 
\[
S^i \asso A\circ \phi_i \asso  S^j \circ \phi_j^{-1} \circ \phi_i =S^j \circ g_{ji} .
\]
As  a consequence,  relying  on the  desired  associative property  of
$\asso$, we should have $S^i \asso S^j \circ g_{ji}$ so that we  can  estimate  the
warping function $g_{ji}$ from a pairwise alignment of $\{S^i,S^j\}$.  
Note that since pairwise alignment of $\{S^i,S^j\}$ does not depend on
the order of the sequences, the proposed estimators satisfy $\hat g_{ji}= \hat g_{ij}^{-1}$. 
Then, we  want to combine  these estimates
$\{\hat  g_{ji} ;  j= 1,\dots , K\}$  to   obtain  an   estimate  of
$\phi_i$. In~\citep{Tang_Muller}, the authors use  the empirical mean
and   notice   that   this    makes   sense   since   $\esp(g_{ji}   |
\phi_i)=\phi_i$ and under the assumption $\esp(\Phi)=Id$~\citep[see   also][]{Ana_Muller}. 
Let us recall that these works come from functional data analysis where the functions $\phi_i$'s are real-valued
functions.   Here, we are dealing  with discrete sequences and   the
empirical mean of a set of paths $g_{ji}$ would not give a proper path
and thus would not correspond to an alignment of $S^i$ to the ancestor
$A$. That is why instead of using empirical mean, we will rather rely on
median values. Note that -- in general -- medians have the advantage of being more
robust than mean values.

From   an   algorithmic  perspective,   a   global  alignment   of
  $X=X_1\cdots X_n$ and
  $Y=Y_1\cdots Y_m$ may be encoded trough the sequence of coordinates $Z_1,\ldots,
  Z_T$,  where  $T$  is  the  length  of  the  alignment,  $Z_1=(0,0),
  Z_T=(n,m)$ and for any $ t= 1,\dots ,  T-1$, we have $Z_{t+1}-Z_t \in \{(1,0),
(0,1), (1,1)\}$.  We let 
\begin{multline}
  \label{eq:Z_space}
  \Z_{n,m}=   \{(Z_1,\ldots,   Z_T) ;  T\ge   1,   Z_1=(0,0),
  Z_T=(n,m), \\
\text{ such that }  Z_{t+1}-Z_t \in \{(1,0),
(0,1), (1,1)\}  \text{ for all } t=1, \dots, T-1\},
\end{multline}
be the set of possible alignments between $X$ and $Y$. The link between the path sequence $Z$ and the warping function
$\phi_{X,Y}$ is as follows
\[
\exists t\ge 1 \text{ such that } Z_t=(k,l) \iff \phi_{X,Y}(k-)=l. 
\] 
Now,  considering  a  median path  boils  down  to
computing median coordinates of points $Z_t$ as explained in the next section.

\subsection{The median procedure}
\label{sec:estim}
In this  section, we explain how  for each sequence  $S^i$, we combine
the set  of pairwise alignments of $S^i$  with $S^j$ ($j=1,\dots , K$) to
obtain an estimate  of the warping function $\phi_i$.  We thus fix a
sequence $S^i$ of length $n_i$ and consider its pairwise alignments with all the other
sequences $S^j$   ($j=1,\dots , K$), including itself (the alignment path is
then the identical function from  $[0,n_i]$ to $[0,n_i]$). We first define
an estimate of the true number $N$ of homologous positions in
our  MSA  (that  is  positions  that were  present  in  the  ancestral
sequence $A$, which are the only positions of $A$ that may be recovered), as a 
median value (constrained to be an integer number) of the set of lengths $\{n_j ;   j=1,\dots , K\}$, namely 
\[
\hat N = \med \{n_j ;   j=1,\dots , K\}.
\]
More precisely, as a convention and  for the rest of this work, we set
the median  of a sequence  of integers $(n_1,\dots,n_K)$  with ordered
values $(n_{(1)},\dots, n_{(K)})$ as follows. Whenever $K$ is even and
the mean value $m$ between the ordered statistics $n_{([K/2])}$ and $n_{([K/2]+1)}$ is not an integer, 
we (arbitrarily) set the median as $\lfloor m \rfloor$.

\begin{figure}[!h]
\centering 

\begin{minipage}{0.25\linewidth}
\begin{tikzpicture}[scale=0.5] 
\draw[->](-0.2,0)--(1.5,0); 
\draw [thin, gray,dotted] (0,0) grid (1,8);
\node(Sui) at (0.5,-0.5) {$S^i_u$}; 
\node(Z1) at (-1.2,2) {\red{$\tilde Z_1(u)$}};
\node(Z2) at (2.2,3.2) {\red{$\tilde Z_2(u)$}};
\draw[thick, blue] (0,1)--(1,1); 
\draw[thick, blue] (0,1)--(1,2); 
\draw[thick, blue] (0,2)--(1,3); 
\draw[thick, blue] (0,3)--(1,3); 
\draw[thick, blue] (0,4)--(1,5);
 \draw[dashed, thick, red] (0,1.9)--(1,2.9); 
\end{tikzpicture}
\end{minipage}
\begin{minipage}{0.45\linewidth}
\begin{tikzpicture}[scale=0.5] 
\draw[->] (0,0)--(0,8.5);
\draw[->] (0,0)--(8.5,0);
\draw [thin, gray,dotted] (0,0) grid (8,8);

\draw[loosely dashed] (8,0)--(8,6);
\draw[loosely dashed] (0,6)--(8,6);
\node (nm) at (9.5,6.5) {$(n_i,\hat N )$}; 
\node (bullet) at (8,6){$\bullet$};

\node (S1) at (0.5,-0.5) {$S^i_1$};   
\node (Sn) at (7.5,-0.5) {$S^i_{n_i}$};   
\node (A1) at (-0.5,0.5) {$A_1$};   
\node (An) at (-0.5,5.5) {$A_{\hat N}$};   

\draw[thick, blue] (0,0)--(1,0)--(3,2)--(3,4)--(4,5)--(8,5);
\draw [thick,blue](0,0) --(8,8);
\draw[thick, blue] (0,0)--(0,2)--(2,4)--(5,4)--(7,6)--(8,6);

\draw[dashed, thick, red] (0.1,-0.1)--(3.1,2.9)--(3.1,3.9)--(4.1,3.9)--(5.1,4.9)--(6.1,4.9)--(7.1,5.9)--(8.1,5.9);

\end{tikzpicture}
\end{minipage}
\caption{On the left: a set of $K=5$ affine vectors (thicked and blue lines)
extracted from  the $K$ alignments of  $S^i $ to the  set of sequences
$S^j$ ($ j=1,\dots, K$), considered  only at  position $S^i_u$.  The
median value is shown  in dashed and red. On the right: a set  of $K=3$ paths (in
blue,  including diagonal path)  standing for  the alignment  of $S^i$
with all other $S^j$ ($ j=1,\dots , K$) and the median path (in dashed and red).}
\label{fig:pos_k}
\end{figure}
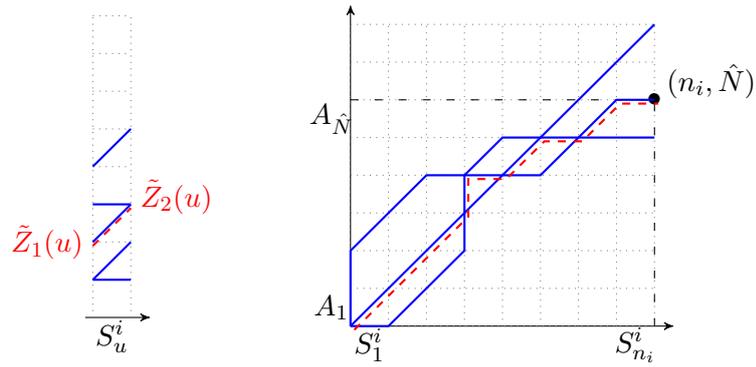

 Now, we aim at defining a path from $[0,n_i]$
to $[0,\hat N]$ by using all the estimated paths $\hat g_{ji}$ obtained
by aligning $\{S^i, S^j\}$. 
For any position $ u =1,\dots ,  n_i$, we consider
the character $S^i_u$ throughout all the pairwise alignments. The
character $S^i_u$ may either be aligned with another character $S^j_v$ or
to a  gap after position  $v$ in sequence  $S^j$, as may be  viewed in
Figure~\ref{fig:pos_k} (left part).
As already explained, each pairwise alignment of $S^i $ with $S^j$
corresponds  to a sequence  of coordinates  $Z^j=(Z_1^j,\ldots, Z_{T_j}^j)
\in \Z_{n_i,n_j}$ as
defined above, see Equation~\eqref{eq:Z_space}. Note that we dropped
the index $i$ on which the sequence $Z^j$ also depends. 
 Now, for each value $ u= 1, \dots, n_i$, there is a unique time point $t_u^j$ and two integers $v_{u,1}^j \le
v_{u,2}^j$ such that
\[
Z_{t_u^j}^j=(u-1,v_{u,1}^j) \text{ and }
Z_{t_u^j+1}^j =(u,v_{u,2}^j).
\]
In practice,  either $v_{u,2}^j=v_{u,1}^j +1$ when  $S^i_u$ is aligned
to a character or  $v_{u,2}^j=v_{u,1}^j$ when  $S^i_u$ is aligned
to a gap. We then compute two median points 
\begin{equation}
  \label{eq:Z_median}
\tilde Z_1(u)= \big( u-1, \med_{ j=1,\dots,  K} \{v_{u,1}^j\} \big)
\quad \text{ and } \quad 
\tilde Z_2(u)=  \big( u, \med_{ j=1,\dots,  K} \{v_{u,2}^j\}
\big) ,
\end{equation}
which  gives  the  part  of  the  median  path  that  concerns  character
$S_u^i$ (Figure~\ref{fig:pos_k}, left part). Note that at this step, we did not take into account the
possible vertical  steps in the alignments $\hat  g_{ji}$. We consider
this now. Indeed, the concatenation of the positions 
\[
(\tilde Z_1(1), \tilde Z_2(1),
\tilde Z_1(2), \tilde Z_2(2), \ldots , \tilde Z_1(n_i), \tilde
Z_2(n_i))
\]
 is almost what we wanted, namely the median path that gives
our  estimate of the  alignment between  $S^i$ and  ancestral sequence
$A$. However one should notice that for each value $1\le u \le n_i-1$, we either
have 
\[
 \tilde Z_2(u) = \tilde Z_1(u+1) \text{ or } \tilde Z_2(u) < \tilde Z_1(u+1) .
\]
In the first case ($ \tilde Z_2(u) = \tilde Z_1(u+1)$), one of the two
points $ \tilde Z_2(u) $ or $ \tilde Z_1(u+1)$ is redundant and should
be removed  to obtain a  proper path in $\Z_{n_i,\hat  N}$. Otherwise,
vertical movements are included in the median
path and we keep both values $ \tilde Z_2(u) $ and $ \tilde Z_1(u+1)$.

Let us note that at the last position $u=n_i$, we have $\tilde Z_2(n_i)=(n_i , \med_{ j=1,\dots,  K} n_j )=(n_i, \hat N)$ so that the
median path correctly ends at $(n_i,\hat N)$.

\subsection{MSA from median warping}
\label{sec:algo}
We  now combine  the different  steps to  obtain our  global alignment
procedure as  explained in pseudocode  in Algorithm~\ref{algo:MSA}. 
This algorithm outputs a set of homologous positions (Hom) as well as a set of inserted positions (Ins). 
Let us recall that in each sequence,  homologous positions correspond to  positions that are
aligned with an ancestor position $A_u$ ($ u=1,\dots , \hat N$). Then two homologous positions may be separated by insert
runs which are not aligned. 
We use a  table  $\Ins$ with  $K$  rows and  $\hat N  +1$
columns. Each entry $(i,u)$ with $ u=2,\dots, \hat N$  
of this table contains the number of insertions in $S^i$ between two homologous
positions  with respective ancestor  positions $A_{u-1}$  and $A_{u}$,
while  entry $(i,1)$  (resp. entry  $(i,\hat  N+1)$) is  the number  of
insertions before first (resp. after last) homologous position. We also use a table $\Hom$ with $K$ rows and $\hat N$ columns. Each entry $(i,u)$
of  this  table  contains  the  position in  sequence  $S^i$  that  is
homologous to  ancestor position $A_u$.  When there is no  such position
(a deletion in sequence $S^i$), the entry is set to $0$.

For instance, going back to the example of a multiple sequence alignment of Section~\ref{sec:MSA}, we may describe this
alignment with the following tables
\[
\Hom =
\begin{pmatrix}
  2 & 3 & 4 & 0\\
  1 &2 & 0 & 6 \\
  2 & 3 &4  & 5
\end{pmatrix}
\text{ and } \Ins = 
\begin{pmatrix}
 1 &0&0&2&1 \\
0 &0&0&3& 0\\
1 &0&0&0& 1  
\end{pmatrix} . 
\]
Note that the information contained in table $\Ins$ is redundant with that in table $\Hom$ but we use it for
convenience. Indeed, it is helpful to output the final alignment whose size (i.e. total number of columns) depends on
the number of homologous positions plus the number of inserted positions. 

Now for each sequence $S^i$ with $i=1,\dots, K$ we reconstruct its alignment with ancestral sequence $A$ as follows. 
For each position $u=1,\dots , n_i$, we compute the coordinates $\tilde Z_1(u), \tilde Z_2(u)$ from
Equation~\eqref{eq:Z_median}. We update the quantity Path($i$) that stores the alignment of sequence $S^i$ to
ancestor $A$ up to position $u$. Then we either update the table $\Ins$ if an inserted position has been obtained
(case $\tilde Z_1(u)[2]=  \tilde Z_2(u)[2]$ where $X[2]$ is the second coordinate of vector $X$) or the table $\Hom$
otherwise.

With respect to the algorithmic complexity of the proposed procedure, we can assume that the $K$ sequences have average
length  similar to $N$. Then, the time complexity of the MSA from median warping (once the pairwise alignments 
are given) is $O(NK^2)$ since we need $K\cdot N$ iterations in which we compute (twice) a median over  $K$ values  (the
complexity of the median calculation being $O(K)$ relying for e.g. on the \texttt{Quickselect}
algorithm~\citep{Floyd}). We provide running time comparisons with \texttt{MAFFT} on the \texttt{BAliBASE} dataset at
the end of Section~\ref{sec:balibase}.

\section{Results}
\label{sec:results}
\subsection{Synthetic experiments}
\label{sec:synthetic}
In this section, we propose a simple synthetic experiment in order to assess the performances of our approach. 
We start with an ancestral sequence $A$ on the set of nucleotides $\A=\{A,C,G,T\}$ with length $N=100$. From this ancestral sequence, we simulate i.i.d. sequences $S^1,\dots, S^K$ as follows. We rely on the simple Thorne-Kishino-Felsenstein~\citep[][hereafter
TKF]{TKF} model that includes an insertion-deletion process  and that can be combined with any substitution
process. Here we set parameters for TKF $\lambda=\mu=0.03$ and use the Jukes-Cantor substitution model~\citep[see for e.g.][]{Yang_book} with
substitution rate $\alpha=0.1$ (all nucleotide frequencies being set to $1/4$). That is, we are simulating $K$
nucleotide sequences related by a star tree and whose branch lengths are set to be equal, as done for instance
in~\citep{Ana_multiplealig}. We repeat this experiment $M=100$ times for each $K$, and we let the number of sequences vary in $\{10, 20, 30, 40, 50\}$. 

To obtain multiple alignments with our procedure, we first conduct all the pairwise alignments between pairs of sequences through the Needleman-Wunsch algorithm for global alignment \citep{Needleman1970} as implemented in the \texttt{pairwiseAlignment} function of the R \texttt{Biostrings} library \citep{Biostrings}. We set the parameters for the pairwise alignment as: gap opening penalty equal to $-10$, gap extension penalty equal to $-0.5$, nucleotide substitution matrix with diagonal values equal to $5$ and non-diagonal values equal to $-4$. These are commonly used as default parameter values for the global alignment of DNA sequences. That is, we do not look at optimising the alignment parameters, but just at showing that our procedure can produce reasonable results under general conditions.

In order to assess the performance of the method, we compute for each alignment two scores that measure its overall concordance with the simulated one, SP (sum of pairs) and TC (total column)
scores (see next section for details), and we compare them to the scores obtained by two well established multiple
alignment softwares, namely \texttt{ClustalW}~\citep{ClustalW} in its \texttt{2.1} version and
\texttt{T-coffee}~\citep{T-coffee} in its \texttt{10.00.r1613} version. Both softwares are run under default parameter values. Finally, we also run our procedure on the reference pairwise alignments, that is, those extracted from the simulated multiple alignment for each pair of sequences. This is done to assess the performance of the method in the best case scenario, as a way to validate the proposed algorithm for combining pairwise alignments.

\begin{figure}
\includegraphics[width=16cm]{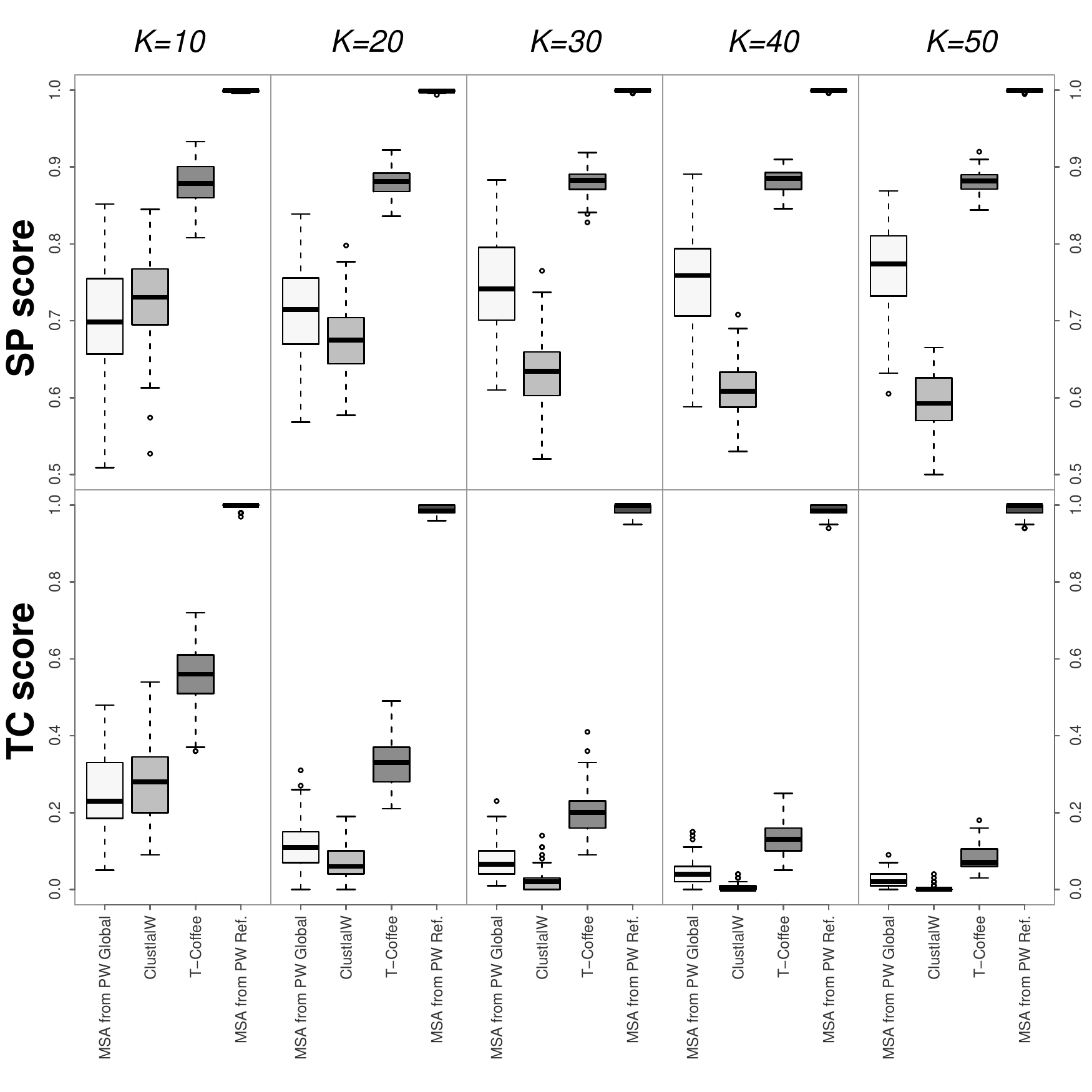}
\caption{SP and TC scores on the synthetic data sets. Distribution over $M$=100 simulated multiple alignments for each value of $K$.}
\label{fig.sim}
\end{figure}

Results are presented in Figure~\ref{fig.sim}. 
The  first thing  to notice  is that  our method,  when used  with the
reference pairwise  alignments, reaches the maximum  score possible in
almost  every case.  Although these  are not  realistic results  since
reference pairwise  alignments are  unknown in practice,  they confirm
the soundness  of our method that  combines pairwise  alignments from
median warping.  With respect to the  other three methods, in  all the
scenarios \texttt{T-coffee} always provides  the best performance.  In
general,  TC scores  tend  to get  worse as  the  number of  sequences
increases whereas SP  scores are more stable.  However, when comparing
our    procedure   (with    estimated    pairwise   alignments)    and
\texttt{ClustalW} (which is a widely used software), we see that \texttt{ClustalW} is better for a small number of sequences, but from $K=20$, the performance of our method is superior.

\subsection{Benchmark results}
\label{sec:balibase}

MSA methods are typically benchmarked on sets of reference alignments, the most widely used being the
\texttt{BAliBASE}~\citep{Balibase}. It is a large scale benchmark specifically designed for multiple sequence
alignment. It  contains test cases based on 3D structural superpositions of protein sequences that are manually refined to ensure the correct
alignment of conserved residues. It is organized into several reference sets, designed to represent real multiple alignment
problems. Reference 1 contains alignments of equidistant sequences with 2 different levels of conservation (RV11 and RV12);   
Reference 2 (RV20) contains families aligned with one or more highly
divergent 'orphan' sequences; Reference 3 (RV 30) contains divergent subfamilies; Reference 4 (RV 40) contains sequences with large
N/C-terminal extensions; and finally  Reference 5 (RV 50) contains sequences with large internal insertions. In addition, three separate
Reference Sets (6–8), are devoted to the particular problems posed by sequences with transmembrane regions, repeats,
and inverted domains~\citep{Balibase}. These last 3 sets are rarely included in benchmark analyses and we exclude them
from ours. For each reference set among the first five ones, except for RV 40, two versions of the same multiple alignments are provided: one with the original sequences (which is noted BB), and one with a shorter version of the sequences contained in the alignment (which is noted BBS). In general, BBS alignments should be easier to recover than the original ones, since the less alignable parts of the original sequences have been removed.

The \texttt{BAliBASE} dataset comes with a function called \texttt{baliscore} used to assess the quality of a MSA. Two
different criteria are used to score an alignment: SP (sum of pairs) and TC (total column)
scores~\citep{Thompson_NAR}. These criteria only use core blocks of the alignment. SP score is the percentage of the
correctly aligned residue pairs in the alignment. It is defined as the number of correctly
aligned residue pairs found in the test alignment divided by the total number of aligned residue pairs in core blocks of
the reference alignment. TC score is the percentage of the correct columns in the alignment. It is defined as the number of correctly aligned columns found in the test alignment divided by the total number of aligned columns in core blocks of the reference alignment.

In order to assess the performance of the method on the Balibase data sets, we proceed as for the simulated alignments
of the preceding  section. The only differences being that we now consider  BLOSUM62 as the default substitution matrix
for pairwise alignments (since we now have protein sequences) and that for the last two reference sets (RV40 and RV50) we consider 'overlap' pairwise alignments instead of global ones. Indeed, these two sets are characterized by large differences in sequences lengths so it is convenient to allow for gaps at the beginning and the end of pairwise alignments. 
In this section we also report the results of \texttt{ClustalW} and \texttt{T-coffee} for reference. We refer to
\citep{Pais2014} for a full comparison of available methods on Balibase. 

\begin{figure}
\hspace*{-1.5cm}\includegraphics[width=18cm]{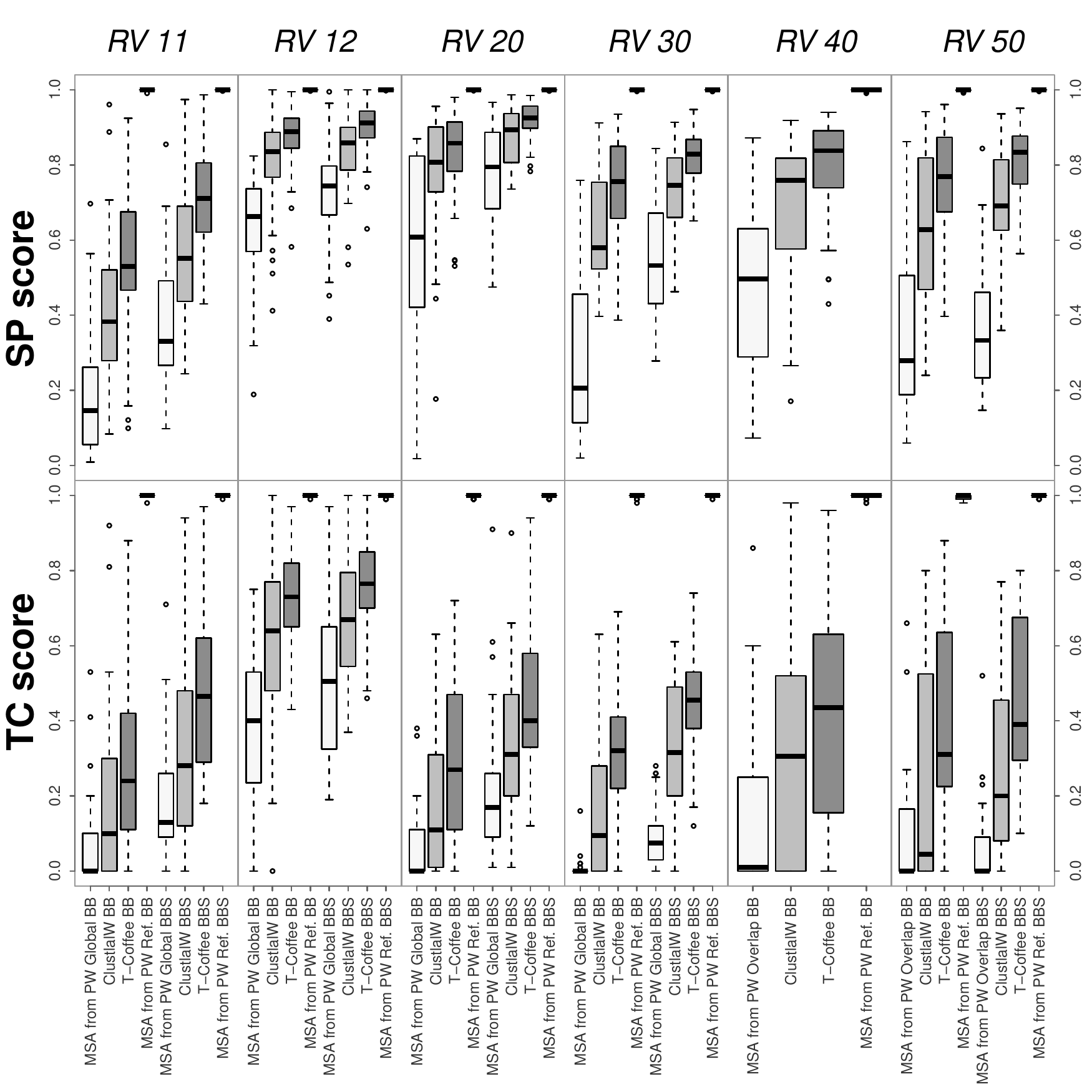}
\caption{SP and TC scores on the Balibase data sets. Distribution over all the multiple alignments of each
reference set. Reference set RV 40 does not provide a short version of the sequences (BBS).}
\label{fig.balibase}
\end{figure}

Figure~\ref{fig.balibase}  shows that  while  our method  is  less
performant  than  \texttt{ClustalW}  and \texttt{T-coffee},  it  still
provides  reasonable   results.  These   results  should  be   put  in
perspective  with  the level  of  complexity  of  the method  that  is
exceptionally low (only  pairwise alignments are needed  as input and
a simple median path is computed). 
Moreover,   as it was  the case  for the
simulated  alignments,  the scores  based  on  the reference  pairwise
alignments are almost always equal to $1$  in all sets for both SP and
TC scores. Again,  these are not realistic results  since the pairwise
alignments extracted  from the  reference multiple alignment  might be
far from the optimal pairwise  alignment between the two sequences for
which   no   extra  information   on   the   remaining  sequences   is
available.  Nevertheless, these  results  are  encouraging since  once
again they  serve to validate our  method from a theoretical  point of
view in the sense that the median warping approach provides sound results.

Finally, let us give an idea of the running times. We ran both \texttt{MAFFT} \citep[in its version 7 with default
values,][]{Mafft}, which is a very fast and widely used tool,  and our procedure on the whole \texttt{BAliBASE} dataset. On 
 a computer with Intel Core i7 at frequency 3.5 GHz and 16 Go RAM (on only one run), we obtained the alignments in 1m23.759s with
\texttt{MAFFT} and 1m32.932s with our procedure. Though our algorithm runs on \texttt{R} with no optimization of any
sort and \texttt{MAFFT} was run through a command line (outside \texttt{R} environment), these running times (which are
not exactly comparable) are roughly the same, which confirms that our procedure is indeed very fast.

\subsection{Conclusions}
\label{sec:ext}
In  this work,  we propose  a proof  of concept  that a  simple method
derived from recent techniques of curve synchronization in the context
of functional data analysis could be of potential interest to MSA. Our
method is able to align a large  number of sequences in a quite simple
and fast manner, although a bit rough w.r.t. accuracy. While we do not
claim to be competitive with actual  aligners,  we   believe  that   our  procedure   could  be
successfully included (for  e.g. as a starting point)  in more refined
MSA techniques. While it is out of  the scope of the current work to
provide such a refined tool, our simulations as well as the use of the
Balibase  dataset  establish  that  the  method  has  good  potential,
particularly when  looking at the  almost perfect results  obtained by
relying on the  (unknown) reference pairwise alignments  (that is, the
pairwise alignments extracted from the  MSA). In particular, while our
experiments rely here on exact scoring alignment with default parameter values,
the method  could be combined  with more refined  pairwise alignments,
such as probabilistic methods that automatically select optimal scoring
parameters for the sequences at stake. 

While  the method  implicitly  assumes that  a  star ultra-metric  tree
describes the evolution  of the sequences from  their common ancestor,
it could be improved to gain  in robustness w.r.t. this assumption. In
particular, let us assume that additional to the sequences one has access to a non
ultra-metric  guide  tree  describing  the common  evolution  of  these
sequences.  
Then, we propose to weight the sequences in a way inversely proportional to their distance to the
  root. Let $d_i$ denote the evolutionary distance (i.e. branch length) from sequence $S^i$ to root and
  $d_{\max}=\max_{i=1,\dots,  K} d_i$. We fix $\epsilon >0$ and set the weight values to 
\[
w_i = \frac {1 - d_i/(d_{\max}+\epsilon)}{\sum_{ k=1,\dots , K} 1-d_k/(d_{\max}+\epsilon)} \in (0,1). 
\]
Then our method could be generalized to the computation of a weighted
median  path, namely  relying  on weighted  median  values instead  of
simple ones. In  such a way, sequences farther from  the ancestor will
have  a   lower  weight  in   the  MSA.   We  leave  this   for  later
investigation.



\newpage 

\newpage

\newpage

\begin{algorithm}
 \CommentSty{// Pairwise alignments Align($i,j$)}\;
For any pair of sequences $\{S^i, S^j\}$, compute their pairwise
alignment Align($i,j$). 
 \BlankLine
 \BlankLine
\CommentSty{// Reconstructing alignment Path($i$) of each $S^i$ with ancestor $A$}\;
$\tilde Z_2(0)$ $\leftarrow ((0,0))$\; 
$\Ins \leftarrow$ table of $0$'s with $K$ rows and $\hat N+1$ columns\;
$\Hom \leftarrow$ table of $0$'s with $K$ rows and $\hat N$ columns\;
\For{$i=1$ to $K$ } 
 {
\CommentSty{// Alignment initialization}\;
Path($i$) $\leftarrow\tilde Z_2(0)$ \\
\For{$u=1$ to $n_i$}
{
Compute the  sequence of median values $\tilde  Z_1(u), \tilde Z_2(u)$
at position $u$ from Equation~\eqref{eq:Z_median}\;
\CommentSty{// Update alignment}\;
\If {$\tilde Z_1(u)=\tilde Z_2(u-1)$} {Path($i$) $\leftarrow$ concatenate
  (Path($i$), $\tilde Z_2(u)$)}
\Else {Path($i$) $\leftarrow$ concatenate
  (Path($i$), $\tilde Z_1(u), \tilde Z_2(u)$)}
\BlankLine
\CommentSty{// Update insertion and homologous positions tables}\;
\If  {$\tilde  Z_1(u)[2]=\tilde  Z_2(u)[2]$} {$\Ins[i,\tilde  Z_1(u)[2]
  +1] \leftarrow \Ins[i,\tilde Z_1(u)[2] +1] + 1$ }
\Else {$\Hom[i,\tilde Z_2(u)[2]] \leftarrow u$}
}
}
\BlankLine
\BlankLine
\CommentSty{// Compute the maximal  number of insertions before first,
  between 2 and after last homologous positions, respectively}\;
NbIns[$\cdot] \leftarrow \max_{ i=1, \dots , K} \Ins[i,\cdot]$
\BlankLine
\CommentSty{// Compute size of multiple alignment}\;
$L \leftarrow \hat N +\sum_{j=1}^{\hat N +1}$ NbIns[$j]$\;
\BlankLine
\BlankLine
 \CommentSty{// Return the multiple alignment}\;
$T \leftarrow$  table with $K$ rows  and $L$ columns  filled with gaps
symbols\;
Insert the  homologous positions from table $\Hom$  at correct positions
in table $T$\;
Insert the inserted positions from table $\Ins$  at correct positions
in table $T$\;
Return(T) \;
\BlankLine
\BlankLine
\caption{Pseudocode for MSA from median warping} 
\label{algo:MSA}
\end{algorithm}

\newpage

\newpage

\end{document}